\documentclass[finalold]{agujournal2019}
\usepackage{url} 
\usepackage{lineno}
\usepackage[inline]{trackchanges} 
\usepackage{soul}
\usepackage{amsmath}
\usepackage{amssymb}
\usepackage{amsthm}
\usepackage{mathrsfs}

\usepackage{xcolor}
\definecolor{darkblue}{rgb}{0,0,.5}
\usepackage[colorlinks=true,
            linkcolor=darkblue, urlcolor=darkblue, citecolor=darkblue,
            raiselinks=true,
            bookmarks=true,
            bookmarksopenlevel=1,
            bookmarksopen=true,
            bookmarksnumbered=true,
            hyperindex=true,
            plainpages=false,
            pdfpagelabels=true,
            pdfstartview=FitH,
            pdfstartpage=1,
            pdfpagelayout=OneColumn]{hyperref}
\usepackage{mathtools}

\usepackage{multirow}

\usepackage{natbib}

\usepackage{array}
\newcolumntype{P}[1]{>{\arraybackslash}p{#1}}
\newcolumntype{M}[1]{>{\centering\arraybackslash}m{#1}}

\usepackage{verbatim}

\usepackage[inline]{trackchanges}
\addeditor{ZN}

\draftfalse

\journalname{the submitted article}

\begin{document}

%
%

\title{Delayed dynamic triggering and enhanced high-frequency seismic radiation due to brittle rock damage in 3D multi-fault rupture simulations}

%
%




\authors{Zihua Niu\affil{1}, Alice-Agnes Gabriel\affil{2,1},
Yehuda Ben-Zion\affil{3,4}}

 \affiliation{1}{Department of Earth and Environmental Sciences, Ludwig-Maximilians-Universit\"at M\"unchen, Munich, Germany}
\affiliation{2}{Scripps Institution of Oceanography, UC San Diego, La Jolla, CA, USA}
\affiliation{3}{Department of Earth Sciences, University of Southern California, Los Angeles, CA, USA}
\affiliation{4}{Statewide California Earthquake Center, University of Southern California, Los Angeles, CA, USA}


\begin{keypoints}
\item We present 3D dynamic rupture models with brittle damage using the discontinuous Galerkin method.
\item Co-seismic off-fault damage generates isotropic high-frequency radiation and modifies rupture speed.
\item We identify a new mechanism for delayed earthquake triggering in fault systems.
\end{keypoints}

\begin{abstract} 
Using a novel high-performance computing implementation of a nonlinear continuum damage breakage model, we explore interactions between 3D co-seismic off-fault damage, seismic radiation, and rupture dynamics. 
Our simulations demonstrate that off-fault damage enhances high-frequency wave radiation above 1~Hz, reduces rupture speed and alters the total kinetic energy.
We identify distinct damage regimes separated by solid-granular transition, with smooth distributions under low damage conditions transitioning to localized, mesh-independent shear bands upon reaching brittle failure. 
The shear band orientations depend systematically on the background stress and agree with analytical predictions. 
The brittle damage inhibits transitions to supershear rupture propagation and the rupture front strain field results in locally reduced damage accumulation during supershear transition.
The dynamically generated damage yields uniform and isotropic ratios of fault-normal to fault-parallel high-frequency ground motions. 
Co-seismic damage zones exhibit depth-dependent width variations, becoming broader near the Earth's surface consistent with field observations, even under uniform stress conditions.
We discover a new delayed dynamic triggering mechanism in multi-fault systems, driven by reductions in elastic moduli and the ensuing stress heterogeneity in 3D tensile fault step-overs. This mechanism affects the static and dynamic stress fields and includes the formation of high shear-traction fronts around localized damage zones.
The brittle damage facilitates rupture cascading across faults, linking delay times directly to damage rheology and fault zone evolution.  
Our results help explain enhanced high-frequency seismic radiation and delayed rupture triggering, improving our understanding of earthquake processes, seismic radiation and fault system interactions.
\end{abstract}

\section*{Plain Language Summary}
Earthquake ruptures perturb the stress state of the surrounding rocks, leading to rock damage with moduli reductions near the rupture zones. 
Based on an advanced nonlinear brittle rheology model and an efficient numerical algorithm, we simulate in 3D dynamic generation of rock damage and how it influences seismic radiation and earthquake source process. 
We identify distinct damage patterns in rocks subjected to damage levels below and beyond their brittle failure threshold. Before the failure points, the damage is spreading smoothly. 
However, once brittle failure occurs, the damage forms localized structures extending from the major fault. We quantify the generated high-frequency motions above 1 Hz due to breaking rocks. This explains components of seismic radiation underrepresented in models ignoring the rapid rock moduli reduction. 
We also discover a new process that can trigger earthquakes on nearby faults with a delay time. This occurs because the weakened rocks create non-uniform stress that can eventually induce slip on another fault at locations with high loads. 
Our findings suggest that off-fault damage plays key roles in rupture dynamics, providing improved ability to understand earthquake processes, near-fault ground motion, and potential triggers for future events.

\section{Introduction}
\label{sec:introduction}
The nonlinear mechanical response of rocks beyond the elastic limit is important for multiple aspects of earthquake rupture dynamics and ground shaking.
Crustal faults are surrounded by hierarchical zones of rock damage with reduced elastic moduli that are generated by and evolve during earthquake ruptures \citep[e.g.,][]{sibson1977fault,chester1993internal,ben2003characterization,mitchell2009nature}.
Off-fault damage alters rupture dynamics by changing the energy partitioning between dissipation and radiation, modifying the seismic wavefield, increasing material and stress heterogeneities, and altering the size of earthquake ruptures and fault interactions \citep{ben2008collective,okubo2019dynamics,johnson2021energy,zhao2024dynamic}. 
However, the co-seismic reduction in elastic moduli is often ignored in theoretical, numerical, and empirical earthquake models. As an example relevant to this study, dynamic reduction of elastic moduli (brittle rock damage) can produce local seismic radiation and stress heterogeneity due to the reduced capacity of damaged rocks to hold the stored elastic strain energy \citep{ben2009seismic,ben2019representation}.

This additional radiation, which is expected to be pronounced around the rupture front and fault segment edges, may facilitate `rupture jumping' producing dynamic triggering of adjacent fault segments.  Off-fault damage may also affect fault system interactions by introducing stress heterogeneity and local bimaterial interfaces \citep{lyakhovsky1997non,sammis2010effects, xu2015dynamic, mia2024rupture}. Previous studies suggest that reduced shear modulus zones promote rupture jumps over larger distances \citep{finzi2012damage} than commonly assumed. 
These effects can lead to larger-than-expected multi-fault earthquakes, with  important implications for seismic hazard assessment.
Earthquake triggering does not always occur at the time of the largest dynamic stress perturbations during the passage of seismic waves \citep[e.g.,][]{yun2024controls}.
Examples include the 2023 Kahranmaras Turkey doublet where a M$_w$ 7.7 earthquake occurred nine hours after a M$_w$ 7.8 event \citep{jia2023complex}, and the 2019 M$_w$ 7.1 Ridgecrest, California, mainshock occurring 34 hours after a M$_w$ 6.4 foreshock \citep{ross2019hierarchical,taufiqurrahman2023dynamics}. 
Other large earthquake pairs have also been separated by minutes to days \citep{hauksson19931992,ryder2012large,sunil2015post}. In this study, we demonstrate that co-seismic non-linear damage processes can contribute to delayed triggering within multi-segment fault systems. 

Brittle damage in earthquake rupture zones incorporating reduction of elastic moduli is not fully captured by commonly used plasticity models. A computationally efficient, high-fidelity approach for modeling these effects in 3D dynamic rupture simulations is currently lacking.
To enable simulations of dynamic ruptures and waves in 3D solids with evolving fault zones, we integrate the nonlinear continuum damage breakage (CDB) model of \citep{lyakhovsky2014continuum, lyakhovsky2016dynamic} into a high-performance discontinuous Galerkin framework. Our optimized implementation makes it feasible to perform large-scale simulations on modern HPC infrastructure of earthquake ruptures with spontaneous generation of brittle damage in regions where the elastic limit has been reached. We demonstrate that this approach captures realistic co-seismic generation of fault damage zones and shear band formation.  We also demonstrate that heterogeneous off-fault moduli reduction can facilitate delayed rupture cascading across faults and produce enhanced isotropic high-frequency radiation beyond 1 Hz.

\section{Methods}
\label{sec:methods} 
We use numerical simulations that extend recent work of \citet{niu2025nlwave} by implementing a Continuum Damage-Breakage (CDB) model \citep{lyakhovsky2014continuum} into 3D dynamic rupture simulations. The CDB model, formulated within continuum mechanics, includes (i) a nonlinear strain energy function of a damaged solid with micro-crack density described by a scalar damage variable ($\alpha$), (ii) an evolution equation for ($\alpha$) based on conservation of energy and non-negative changes of entropy, and (iii) a transition at a critical $\alpha$ to dynamic instability and a granular phase described by a breakage variable ($B$) for post-failure grain size distribution 
\citep{lyakhovsky1997distributed,einav2007breakage1,einav2007breakage2,lyakhovsky2014continuum,lyakhovsky2016dynamic}. This phase transition avoids the non-convexity of the solid phase at large damage \citep{lyakhovsky2014continuum}. Physically, it enables the CDB model to capture additional high-frequency radiation emanating from the damaging off-fault material  \citep{ostermeijer2022evolution}.

We solve the governing equations using a discontinuous Galerkin method in the open-source code SeisSol \citep{uphoff_2024_14051105}. The stress-strain relationships for the pre-failure solid and post-failure granular phases of rocks are represented with the two material state variables $\alpha$ and $B$ \citep{lyakhovsky2014continuum, lyakhovsky2016dynamic}, which evolve in time through a nonlinear system of conservation laws as functions of strain invariants $r_{\alpha}$ and $r_B$ detailed in the SI. We use a face-aligned coordinate transformation for accurate stress estimation at frictional interfaces \citep{pelties2012three}, integrating dynamic rupture with various friction laws \citep{uphoff2020flexible}.
To efficiently resolve nonlinear wave interactions and co-seismic damage in 3D, we employ a parallelized MPI/OpenMP implementation for high-performance computing. Additional methodological details, including full equations and numerical implementation, are provided in the SI.

\section{Results}  
\label{sec:results_1}

We systematically investigate how co-seismic off-fault damage influences 3D dynamic rupture, near-fault seismic radiation, and fault system interaction, focusing on three key aspects:
(1) the evolution of off-fault rock damage and energy radiation before and beyond the solid-granular phase transition (Sec. \ref{subsec:31}), (2) the role of off-fault energy dissipation in modulating rupture dynamics, including supershear transition (Sec. \ref{subsec:32}), and (3) the effects of co-seismic off-fault damage on earthquake interaction within a multi-fault system (Sec. \ref{subsec:33}).

\subsection{Two end-members of co-seismic off-fault damage}
\label{subsec:31}

We use the dynamic rupture community benchmark problem TPV3 \citep{harris2009scec}, which features a right-lateral vertical strike-slip fault in a half-space. Our 3D domain spans 120 km $\times$ 120 km $\times$ 60 km, with a 30 km long, 15 km deep fault governed by a linear slip-weakening friction law \citep{ida_1972, Palmer_and_Rice_1973,Andrews_1976,Day_1982}. Additional material properties and initial background stresses required to extend the benchmark setup to non-linear CDB damage rheology are listed in Table S1. Among the parameters in the CDB model, the damage evolution coefficient $C_d$ in Eq. (2) of the SI controls the damage levels in off-fault rocks.

We examine two end-member cases: (1) \textit{small} co-seismic damage ($C_d$ = 5 $\times$ 10$^{-6}$ (Pa$\cdot$s)$^{-1}$), where the bulk rock remains in the solid regime, versus (2) \textit{large} co-seismic damage ($C_d$ = 6 $\times$ 10$^{-5}$ (Pa$\cdot$s)$^{-1}$), where off-fault rocks close to the rupture front transition to a granular state within ~0.01 s.

\begin{figure}[hptb]
    \centering
    \includegraphics[width=1\columnwidth]{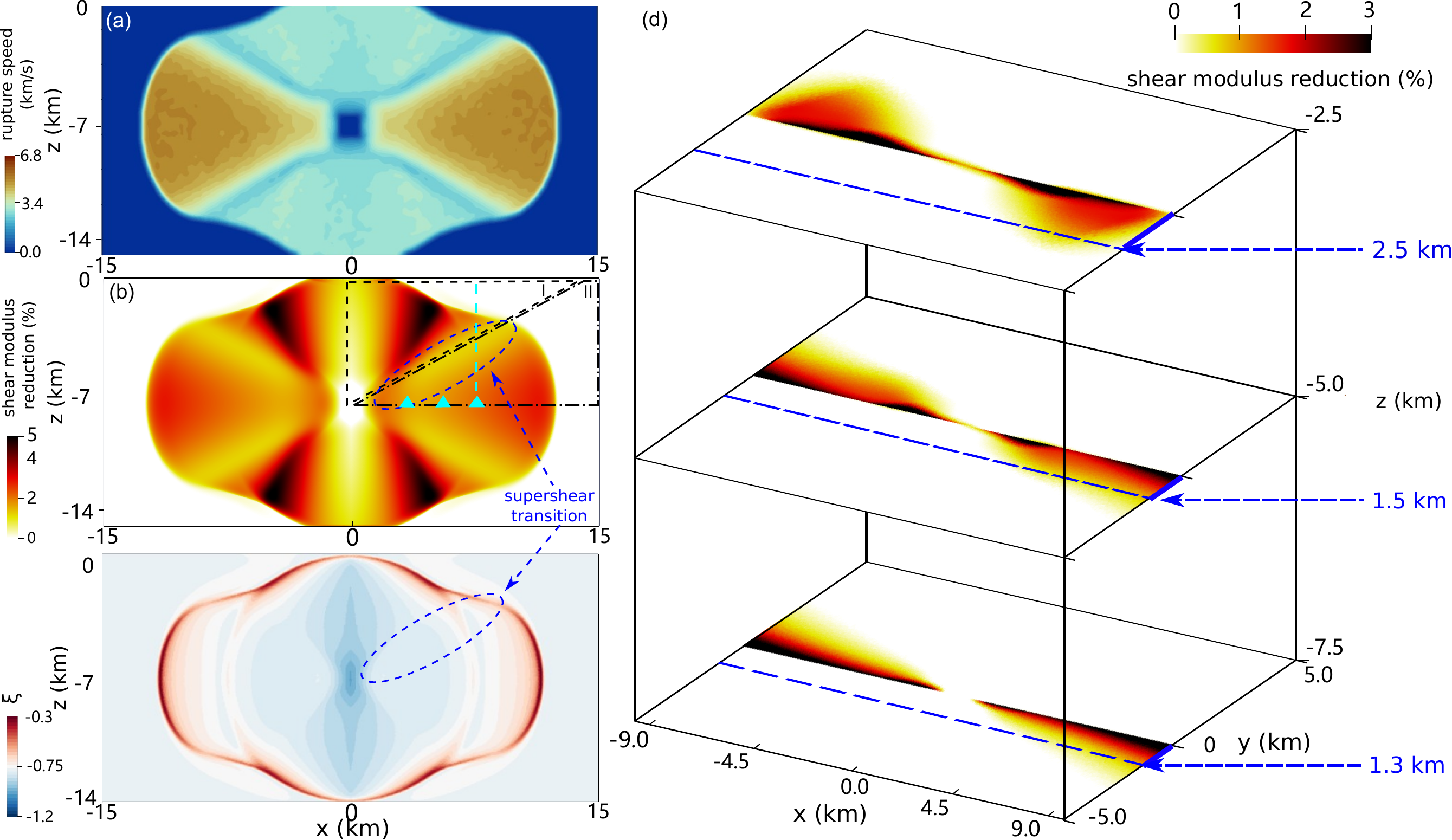}
    \caption{3D rupture dynamics with small off-fault damage that remains below the threshold for solid-granular phase transition. (a) Distribution of rupture speed on the fault plane 2.5~s after rupture onset. The supershear region (rupture speed $\ge$ shear wave speed, 3.4~km/s) is highlighted in red. (b) Shear modulus reduction in off-fault material next to the fault plane. The sub-Rayleigh (I) and supershear rupture (II) regions are marked, respectively, in dashed and dash-dotted black curves. The location of supershear transition is marked as a dashed blue circle. (c) Distribution of the strain ratio $\xi$ at 2.5~s in the bulk material next to the fault. (d) Cross-fault damage distribution at 7.5~km, 5.0~km, and 2.5~km depths, illustrating depth-dependent variations in damage patterns.}
    \label{Fig1:small damage, at different depths}
\end{figure}

For the small damage case, Fig. \ref{Fig1:small damage, at different depths} illustrates the off-fault damage distribution 2.5 s after rupture onset and its effect on dynamic rupture.
The chosen background stress and model parameters lead to bilateral along-strike supershear transitions (from blue to red regions, Fig. \ref{Fig1:small damage, at different depths}a) as a result of a daughter crack that nucleates in front of the sub-Rayleigh rupture due to the local dynamic stress peak \citep{Andrews_1976,dunham2007conditions}. This contributes to the complex off-fault damage distribution (Fig. \ref{Fig1:small damage, at different depths}b). As indicated in Fig. \ref{Fig1:small damage, at different depths}b, we categorize off-fault damage into two regions based on the rupture speed:
Region I associated with a sub-Rayleigh rupture speed and Region II with a supershear rupture speed. The largest fault zone shear modulus reduction (up to 5\%) occurs within Region I, while in Region II it remains below 3\%. In particular, the modulus reduction is lower than 1\% around the supershear transition region (circled in blue). 

The modeled damage level is highly dependent on the shape of the strain tensor in rocks close to the fault surface. In the CDB model, this is parameterized as $\xi = I_1/\sqrt{I_2}$ according to Eq. (2) in the SI, where $I_1$ and $I_2$ are the first and second strain invariants. We show the distribution of $\xi$ around the fault plane in Fig. \ref{Fig1:small damage, at different depths}c. The regions with a higher strain ratio ($\xi \approx -0.3$, in red) at the rupture front correspond to regions with greater shear modulus reduction in Fig. \ref{Fig1:small damage, at different depths}b. Within the supershear transition zone, we observe a lower strain ratio ($\xi \approx -0.6$) around the rupture front. This contributes to locally weaker damage. Conversely, regions with $\xi < -0.75$ (in blue) accumulate zero damage as a consequence of the imposed model parameter $\xi_0 = -0.75$ in Table S1, which is chosen following \citet{lyakhovsky2016dynamic} and corresponds to an internal friction angle of 43$^{\circ}$ in the Mohr-Coulomb failure criterion of rocks \citep{griffiths1990failure}. We show how the supershear transition leads to a lower $\xi$ at the rupture front and influences the accumulation of damage in Movie S1.

In addition to along-strike variations, we observe a pronounced depth-dependence of off-fault damage (Fig. \ref{Fig1:small damage, at different depths}d, Movie S1). At 2.5 km, the damage zone with a shear modulus reduction greater than 1\% extends laterally to $\sim$2.5 km) from the fault, whereas it remains more localized ($\sim$1.3 km) at 7.5 km depth. 
Field studies provide observational support for this result, consistently documenting damage zones that systematically narrow with increasing depth \citep[e.g.,][]{sylvester1988strike,faulkner2011scaling,ben2019spatial}.
Previous 2D and 3D simulations show such a flower-like depth-dependent fault zone width as a result of lower confining stress at shallower depths \citep{ben2005dynamic,ma2010inelastic,okubo2019dynamics,ferry2025depth}. Due to higher peak slip rates at shallower depths (Fig. \ref{Fig1-1:small damage, receivers}b), the presented 3D simulations with the CDB model indicate that such flower-like off-fault damage may also emerge under a uniform background stress.

\begin{figure}[hptb]
    \centering
    \includegraphics[width=0.5\columnwidth]{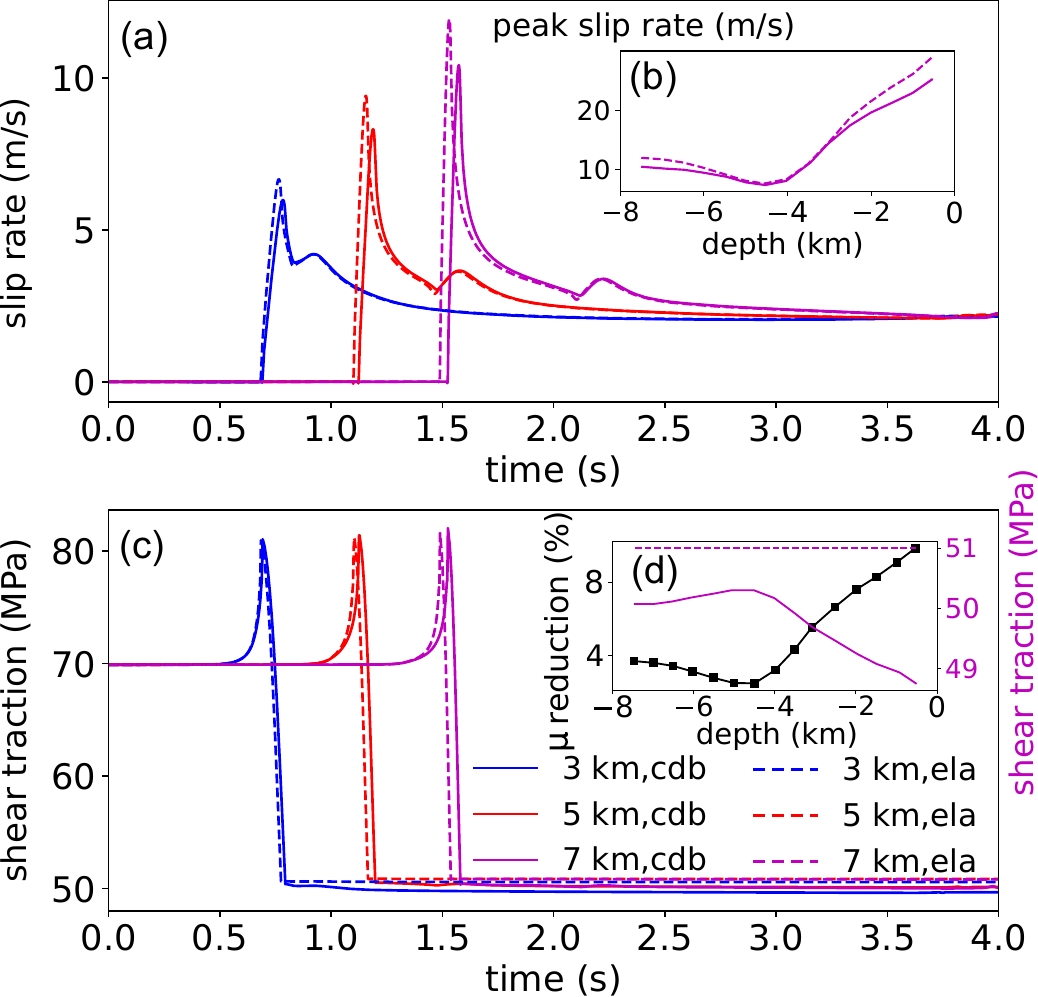}
    \caption{\small Comparison between elastic and CDB models with off-fault damage below the solid-granular phase transition threshold. (a) Slip rate time series at three on-fault receivers (cyan rectangles in Fig. \ref{Fig1:small damage, at different depths}) located at $x$ = 3, 5, and 7 km. Dashed curves represent the purely elastic off-fault material reference simulations, whereas solid curves correspond to simulations incorporating non-linear off-fault damage simulations with the CDB model. (b) Variation of peak slip rate with depth along a cross-section indicated by the dashed cyan line in Fig. \ref{Fig1:small damage, at different depths}. (c) Shear traction time series at the same three on-fault receivers as in (a). (d) Depth profile of post-rupture shear traction and shear modulus ($\mu$) reduction along the dashed gray survey line in Fig. \ref{Fig1:small damage, at different depths}. Note the inverse correlation between shear modulus reduction and post-rupture shear traction.}
    \label{Fig1-1:small damage, receivers}
\end{figure}

In Fig. \ref{Fig1-1:small damage, receivers}, we compare the slip rate, shear traction, and damage accumulation at three receivers (cyan triangles) in Fig. \ref{Fig1:small damage, at different depths}b between the CDB model and the linear elastic model.
Rupture speed decreases by 4\% due to energy dissipation in the generation of off-fault damage as indicated in the time series of the slip rate (Fig. \ref{Fig1-1:small damage, receivers}a). This effect also results in up to 12\% lower peak slip rates 7 km away from the nucleation center compared to the case with elastic off-fault model (dashed curves in Fig. \ref{Fig1-1:small damage, receivers}a).
These 3D results are consistent with previous 2D dynamic rupture simulations with off-fault damage \citep{xu2015dynamic} or incorporating elastoplasticity \cite{andrews2005,wollherr2018off}. 
Analysis of peak slip rates (Fig. \ref{Fig1-1:small damage, receivers}c) along a cross-section that connects Region I with Region II (the dashed gray line in Fig. \ref{Fig1:small damage, at different depths}b), shows the lowest peak slip rate inside the supershear transition region.
Comparing the elastic reference model and the CDB model, the largest difference ($\sim$13\%) in peak slip rate occurs at the free surface, highlighting pronounced near-surface weakening.

Additionally, post-rupture shear traction is notably lower in damaged regions (Fig. \ref{Fig1-1:small damage, receivers}c), particularly in areas experiencing the largest shear modulus reduction (Fig. \ref{Fig1-1:small damage, receivers}d). The highest modulus reduction and associated traction drop coincide within the supershear transition zone.
Along the cross-section indicated in Fig. \ref{Fig1:small damage, at different depths}b, post-rupture shear traction remains constant at 51~MPa in the elastic model (Fig. \ref{Fig1-1:small damage, receivers}d). In contrast, simulations including non-linear off-fault damage (CDB model) show post-rupture traction variations between 48.7 MPa and 50.2 MPa, with the maximum traction observed within the supershear transition region.

Under conditions where damage approaches the solid-to-granular transition threshold within the CDB framework, the stress-strain relationship will rapidly change from the solid type, that is, $B = 0$ in Eq. (2) in the SI, to the granular type, that is, $B = 1$. This transition leads to highly localized deformation that forms off-fault shear bands. In this state, the off-fault damage pattern differs markedly from the more distributed damage observed at lower levels.

\begin{figure}[hptb]
    \centering
    \includegraphics[width=0.85\columnwidth]{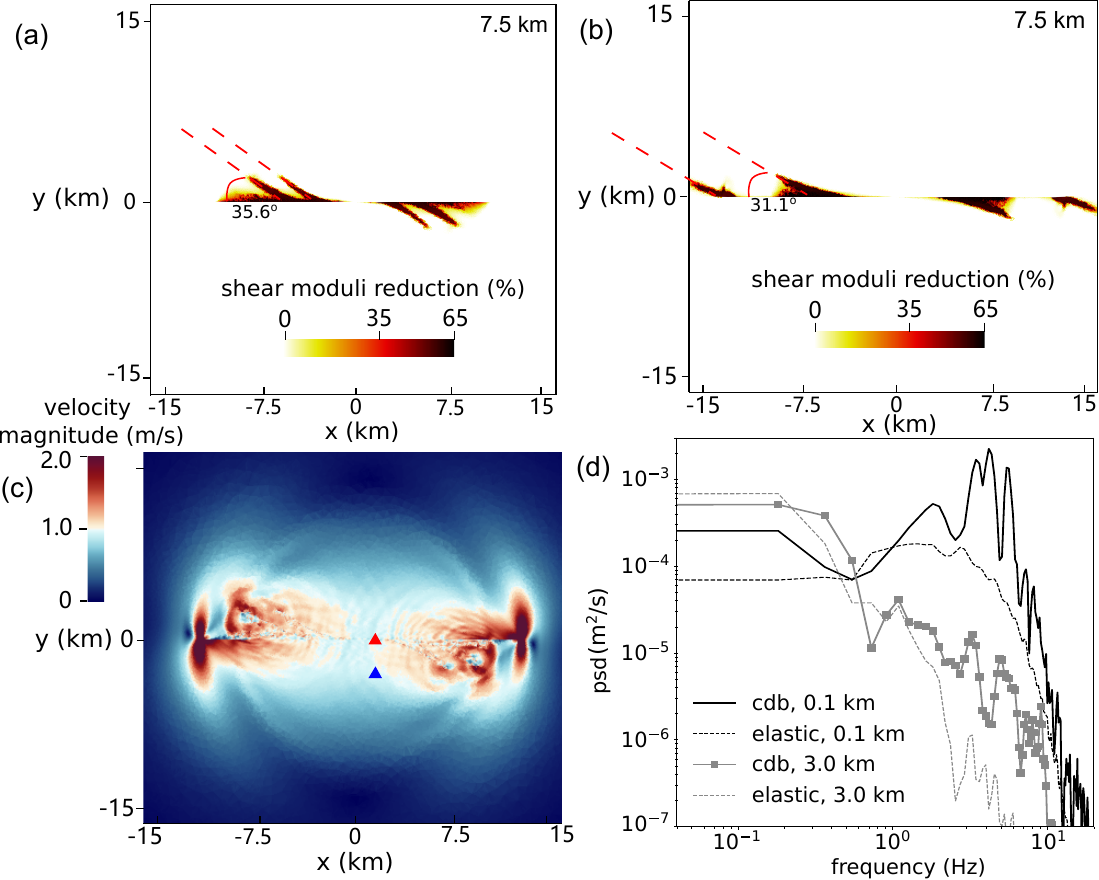}
    \caption{\small CDB 3D dynamic rupture simulations with damage level reaching the solid-granular phase transition. The damage distributions for maximum compressive stress oriented 59.1$^{\circ}$ and 54.6$^{\circ}$ from the $x$-axis at the depth of 7.5 km, 3 s after the rupture onset are, respectively, shown in (a) and (b). 
    (c) illustrates the velocity magnitude distribution at 7.5~km depth corresponding to the scenario in panel (a), highlighting two receiver locations marked by the red rectangle at (1.0, -0.1)~km (R1) and the blue rectangle at (1.0, -3.0)~km (R2).
    Panel (d) compares the power spectral density (PSD) of seismograms recorded at these receivers (solid curves) against those obtained from simulations with linear elastic off-fault material (dashed curves), emphasizing the influence of nonlinear damage on seismic wavefield characteristics.}
    \label{Fig2:large damage, different angles, depths}
\end{figure}

Fig. \ref{Fig2:large damage, different angles, depths}a  shows the off-fault damage distribution at a depth of 7.5 km for a maximum compressive principal stress oriented 59.1$^{\circ}$ relative to the fault plane. Under this background stress orientation, distinct shear bands form extending from the fault into the non-linearly deforming off-fault material at an angle of $\sim$35.6$^{\circ}$. This is consistent with analytical predictions based on the CDB model (parameters detailed in Table S1), verifying our approach. We detail how the results from numerical simulations compare to analytical solutions in Text S4 of the SI. To confirm the robustness of the achieved agreement, we vary the orientation of the maximum compressive principal stress towards the fault plane from 59.1$^{\circ}$ to 54.6$^{\circ}$ (Fig. \ref{Fig2:large damage, different angles, depths}b). Correspondingly, the shear bands form at a smaller angle ($\sim$31.1$^{\circ}$) to the fault, maintaining close alignment with the analytical predictions \citep{lyakhovsky1997distributed}. Importantly, the simulated damage patterns remain stable and consistent under mesh refinement from 100 m to 25 m, confirming mesh independence (Fig. S1). The mesh independence is essential to ensure the reliability of the modeled interactions between rupture dynamics and off-fault damage accumulation. We discuss this in more detail in \ref{app:numerics}.

The co-seismically evolving, localized off-fault shear bands generate high-frequency seismic waves. Fig. \ref{Fig2:large damage, different angles, depths}c shows the secondary wave field generated in regions where the solid-granular phase transition occurs. 
We show how these transitions alter the frequency characteristics of seismograms at two receivers in a different way from the linear elastic scenario shown in Fig. \ref{Fig2:large damage, different angles, depths}d. 
At both locations, frequencies between 2 and 5 Hz are enhanced by the secondary wave field, with larger enhancement closer to the fault.

\begin{figure}[hptb]
    \centering
    \includegraphics[width=0.8\columnwidth]{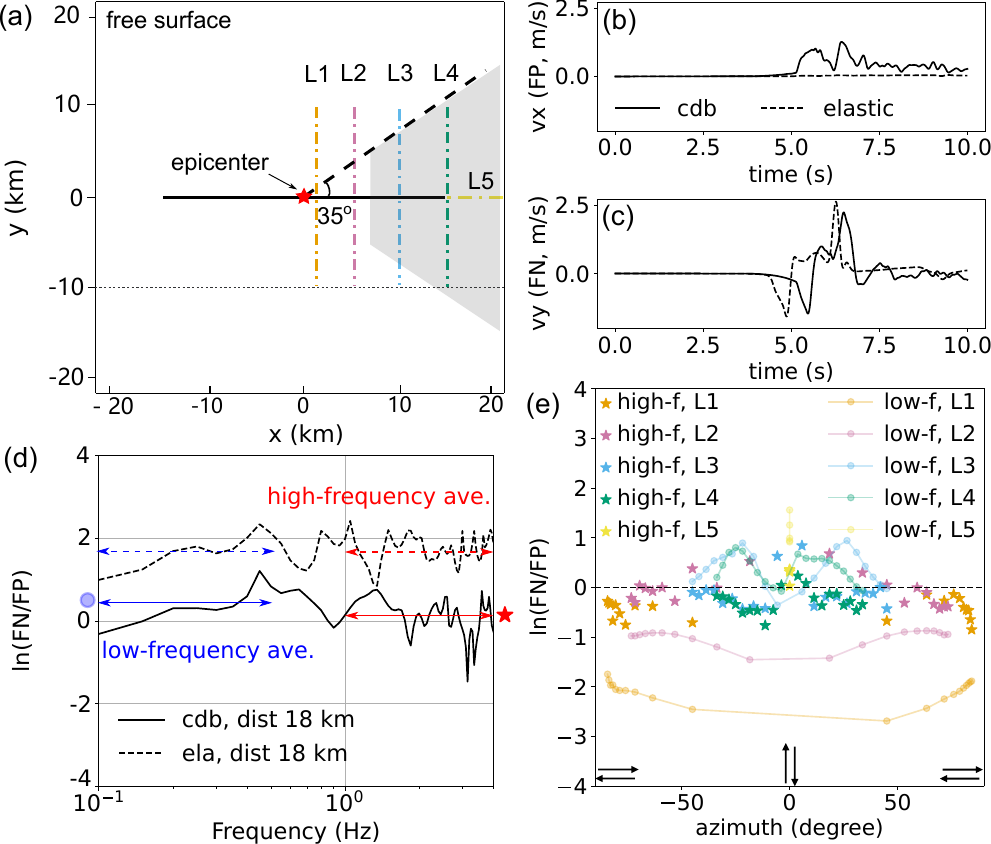}
    \caption{\small Fault-normal (FN) and fault-parallel (FP) ground motions close to the dynamic rupture fault plane. (a) Survey lines on the free surface located at distances $x$ = 1 km (L1), 5 km (L2), 10 km (L3), and 15 km (L4) perpendicular to the fault (solid black line), and along $y$ = 0 km (L5), parallel to the fault. The shaded gray area indicates the region where FN ground motions are expected to exceed FP ground motions for a pure shear (double-couple) source \citep{ben2024isotropic}. 
    Examples of FP (b) and FN (c) ground motions generated by elastic (dashed curves) and CDB non-linear damage (solid curves) simulations recorded at one receiver along L5, located at ($x$,$y$) = (18,0)~km. (d) $\ln{(\text{FN}/\text{FP}})$ frequency amplitude spectra computed at the receiver shown in (b) and (c). Average values within a low-frequency band of [0.1,0.5] Hz (low-f) and a high-frequency band [1,4] Hz (high-f) are highlighted by blue and red arrows, with a circle and a star, respectively. 
    (e) Variations of $\ln{(\text{FN}/\text{FP}})$  ratios from the CDB simulation with azimuth angle along different survey lines indicated in (a). Circles and stars represent low-frequency and high-frequency band averages, respectively. Each marker corresponds to one receiver in (a) and the marker colors in (e) match the line colors in (a).}
    \label{Fig2-1:fnfp_ground_motions}
\end{figure}

Analytical results indicate that damage generation should produce high frequency radiation with significant isotropic component \citep{ben2009seismic,ben2019representation}. To check if this is the case for the enhanced high frequency radiation in the CDB simulation, we examine in Fig. \ref{Fig2-1:fnfp_ground_motions} the variability of the fault-normal (FN) and fault-parallel (FP) ground motions at varying frequencies and receiver locations. 
Receivers placed every 1 km along five survey lines shown in Fig. \ref{Fig2-1:fnfp_ground_motions}a enable a detailed assessment of ground-motion characteristics.  
Figs. \ref{Fig2-1:fnfp_ground_motions}b,c display ground velocities at a receiver located 18 km from the hypocenter along the survey line L5 in Fig. \ref{Fig2-1:fnfp_ground_motions}a. The results 
demonstrate that the dynamic generation of off-fault damage reduces the difference between FN and FP ground motion amplitudes relative to the elastic case. The FP component is almost zero in the elastic case, while the CDB simulation including off-fault modulus reduction produces a more isotropic wavefield with significant FP motion.

In Fig. \ref{Fig2-1:fnfp_ground_motions}d the frequency amplitude spectra of the logarithmic ratio between FN and FP ground motions, referred to as $\ln{(\text{FN}/\text{FP}})$, are shown at the same receiver for low-frequency (0.1 to 0.5 Hz) and high-frequency (1 to 4 Hz) components of ground motions. In the elastic simulation, the logarithmic ratio $\ln{(\text{FN}/\text{FP}})$ is approximately 1.8 for both the low-frequency (blue dashed line) and high-frequency (red dashed line) bands, as expected for a radiation pattern dominated by a pure shear source. In contrast, the CDB simulation produces significantly lower ratios and a transition to radiation that is approximately isotropic at high frequencies. The simulated $\ln{(\text{FN}/\text{FP}})$ is $\sim$0.2 between 0.1 and 0.5 Hz and nearly zero (i.e., FP $\approx$ FN) for high frequencies between 1.0 and 4.0 Hz. The simulated pattern for the CDB results is similar to observed $\ln{(\text{FN}/\text{FP}})$ ratios near earthquake rupture zones \citep{graves2016kinematic,ben2024isotropic}.

To investigate more systematically the amplitudes of FN and FP ground motions in the CBD model, Fig. \ref{Fig2-1:fnfp_ground_motions}e presents results at different locations and frequency ranges. We calculate $\ln{(\text{FN}/\text{FP}})$ at all receivers along the five survey lines in Fig.\ref{Fig2-1:fnfp_ground_motions}a and examine the azimuthal dependence of the ratios. Within the low-frequency band (circles), the FN components are smaller than FP ($\ln{(\text{FN}/\text{FP}})$ $<$ 0) along the survey lines L1 and L2, but exceed FP ($\ln{(\text{FN}/\text{FP}})$ $>$ 0)  along lines L3, L4 and L5, consistent overall with shear dominated S-wave radiation patterns \citep{aki2002quantitative}. In contrast, at high frequencies (stars), $\ln{(\text{FN}/\text{FP}})$ remains close to zero (FN $\approx$ FP), indicating a more isotropic wavefield and a reduced dependence on azimuth. The results show that the co-seismic rock damage leads to a combined shear and volumetric radiation with near-homogeneous isotropic ground motions at higher frequencies.

\subsection{Damage-induced off-fault energy dissipation}
\label{subsec:32}

As shown above, the rapid modulus reduction associated with damage formation produces additional high-frequency seismic radiation, thereby impacting both rupture dynamics and near-fault ground motions. Concurrently, the strain energy stored in the surrounding rock volume is also partially dissipated through the modulus reduction, altering the energy budget of the earthquake.
Earthquake rupture dynamics, such as its propagation speed, size, and interaction across fault systems, which determine an earthquake’s potential impact, are directly related to the nature and amount of energy dissipation involved in the rupture process \citep{shi2009slip,kammer2024earthquake,gabriel2024fault}.

\begin{figure}[hptb]
    \centering
    \includegraphics[width=1.0\columnwidth]{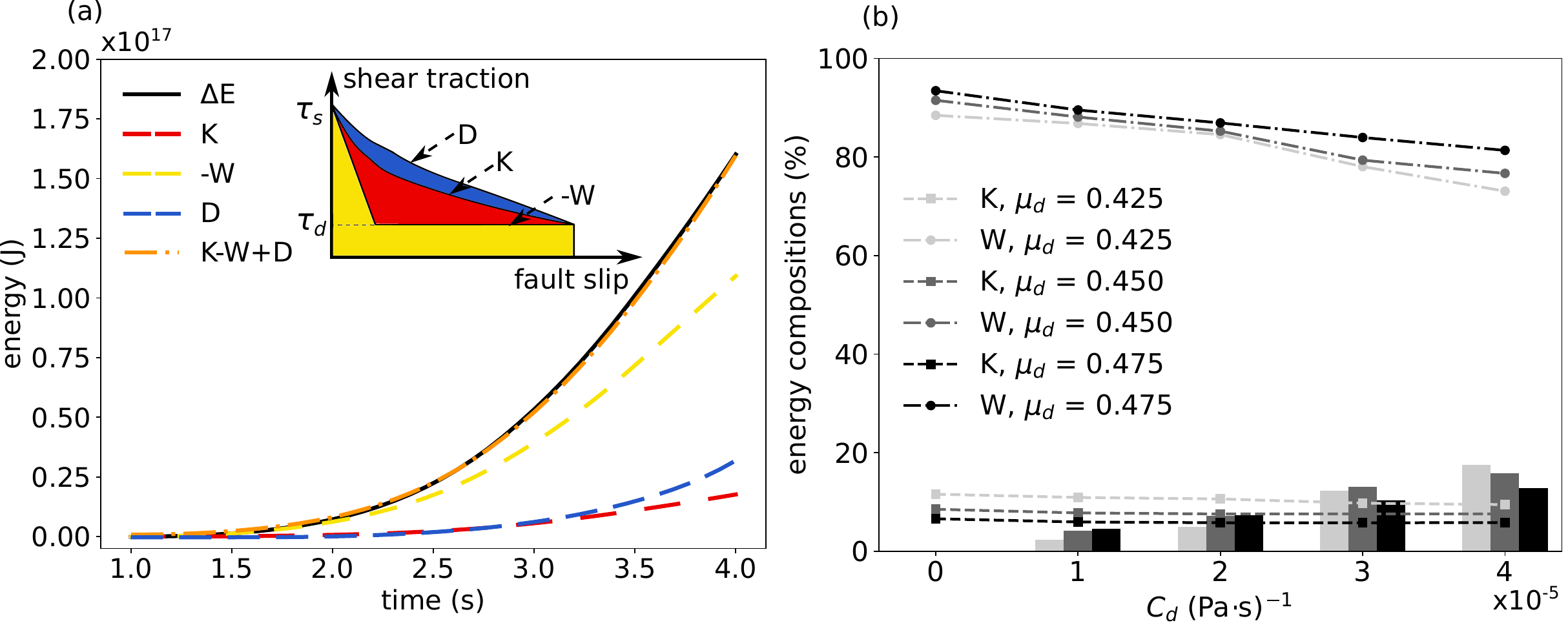}
    \caption{\small Energy budget of CDB dynamic rupture simulations with co-seismic non-linear off-fault damage. (a) Temporal evolution of energy components during rupture propagation. The dashed red curve shows the radiated kinetic energy ($K$), and the dashed yellow curve denotes frictional work on the fault ($-W$). The dashed blue curve represents energy dissipated by off-fault damage evolution ($D$). The inset illustrates the balance of energies during fault slip. 
    (b) Proportions of energy components at the time when the rupture reaches the fault boundary. Dashed lines represent radiated kinetic energy ($K$), dash-dotted lines indicate frictional energy dissipation ($W$), and bars show the percentage of energy dissipated by off-fault damage for varying damage evolution coefficients ($C_d$). The initial stress conditions are identical to those in Fig.~\ref{Fig2:large damage, different angles, depths}a, and model parameters are provided in Table S1.}
    \label{Fig3:energy budget}
\end{figure}

We verify that our simulations accurately conserve energy, that is, the independently computed energy components (Text S3) are evolving consistently with energy conservation laws. 
The energy driving rupture dynamics originates from the drop in stored mechanical potential energy $\Delta E$ in the bulk rock material
defined in Eq. (17) of the SI).
Similarly to the elastic case, this energy is primarily partitioned into frictional work ($-W$) along the fault and radiated kinetic energy ($K$). However, in the CDB model, an additional portion of energy is dissipated through co-seismic off-fault damage generation ($D$, Eq.(9) of the SI), increasing the crack density and the entropy of the system. 
Each of these components accumulates over time (Fig.\ref{Fig3:energy budget}a), and the sum $K - W + D$ closely matches the released mechanical potential energy $\Delta E$, explicitly verifying energy conservation.

Non-linear off-fault energy dissipation significantly delays or inhibits the transition from sub-Rayleigh to supershear rupture speeds. 
A systematic relationship between increased damage evolution coefficient ($C_d$) and delayed supershear transition is illustrated in Fig.~S3.
Energy dissipated in off-fault regions reduces rupture speed, resulting in a larger cohesive zone size along strike compared to the elastic model (Fig.~S4). The slower rupture propagation leads to lower shear traction ahead of the rupture front, impeding the onset of intersonic (supershear) speeds \citep{dunham2007conditions}. At a frictional strength excess to maximum possible stress drop ratio $S$ \citep{Andrews_1976} of 0.6 (Eq. (18) in the SI), the distance between the location of supershear transition and the nucleation center in the along strike direction is $\sim$10\%, $\sim$30\%, and $\sim$120\% longer than the distance in the elastic case, respectively, for $C_d=$ $1 \times 10^{-5}$, $2 \times 10^{-5}$, and $3 \times 10^{-5}$ $(\text{ Pa}\cdot\text{s})^{-1}$.
An increased cohesive zone size has been reported in simulations involving discrete off-fault fracture networks \citep{okubo2019dynamics} and elastoplastic off-fault deformation \citep{wollherr2018off}, the latter also affecting supershear transition \citep{gabriel2013source}. For example, at an $S$ ratio of $S=0.6$, the propagation distance required to transition to supershear speed in 2D simulations with off-fault plasticity by \citet{gabriel2013source} is $\sim$60\% longer than for the elastic case. This increase is comparable  to our simulations with the CDB model using a damage evolution coefficient $C_d$ between $2 \times 10^{-5}$ and $3 \times 10^{-5}$ $(\text{ Pa}\cdot\text{s})^{-1}$.

Increasing off-fault damage systematically shifts energy dissipation from fault friction into the surrounding rock, affecting the earthquake energy budget.
In Fig. \ref{Fig3:energy budget}b, we show how the proportion of frictional energy dissipation decreases consistently with increasing damage evolution coefficient ($C_d$) across all examined dynamic friction coefficients ($\mu_d$). Notably, frictional dissipation decreases more rapidly at lower values of $\mu_d$.
Consequently, at the largest explored damage evolution coefficient ($C_d = 4 \times 10^{-5}$), the proportion of off-fault energy dissipation (bar plots in Fig.\ref{Fig3:energy budget}b) is lowest for the highest friction coefficient ($\mu_d = 0.475$), indicating that stronger frictional resistance limits energy dissipation in the surrounding rock.
The maximum off-fault energy dissipation reaches approximately 17\%, roughly four times larger than the maximum proportion of off-fault fracture energy reported by \citet{okubo2019dynamics}. 
This difference may arise from two reasons. First, their fracture energy calculation does not include frictional heating from discrete fractures. When accounting for this frictional heating, which is roughly four times greater than their reported fracture energy, the total off-fault energy dissipation in their discrete fracture simulations may align closely with our continuum-based CDB model results.
Second, their discrete representation of off-fault fractures may underestimate the energy dissipation in elements that are not predefined by the mesh as potential weak planes able to host failure.
Although off-fault energy dissipation competes directly with on-fault frictional work, the proportion of radiated kinetic energy $K$ remains largely unchanged as off-fault damage increases (higher $C_d$ values). The damped kinematic energy in producing off-fault damage is in part compensated by the additional high-frequency radiation during the rapid solid-granular phase transition. The generated $K$ is primarily controlled by the dynamic friction coefficient $\mu_d$, decreasing from $\sim$10\% for $\mu_d=0.425$ to $\sim$6\% for $\mu_d=0.475$. This suggests that off-fault damage minimally affects the \textit{dynamic} stress amplitudes. This result is in stark contrast to the impact on the \textit{static} stress field, which we will examine in the next Section. 

\subsection{Delayed dynamic triggering facilitated by co-seismic off-fault damage}
\label{subsec:33}

We identify a previously unrecognized mechanism whereby localized off-fault damage introduces sufficient stress heterogeneity to enable delayed dynamic triggering across geometrically disconnected fault segments.
Co-seismic reduction in rock moduli within off-fault shear bands induces static stress heterogeneities influencing the 3D interaction of the fault system.
Laboratory experiments demonstrate a significant rock modulus reduction associated with increasing damage levels at high stress \citep{lockner1977changes,hamiel2009brittle},
an effect not fully captured by elastic or simpler plasticity models.
The realistic modulus reduction in our 3D simulations illustrates how stress heterogeneity generated by localized off-fault damage facilitates delayed dynamic triggering across step-over fault geometries.

To investigate this delayed triggering mechanism, we employ a 3D two-fault model setup from the TPV23 community benchmark \citep{harris2018scec}. 
Compared to the simpler, single strike-slip fault setup (TPV3) in Secs. \ref{subsec:31} and \ref{subsec:32}, TPV23 employs the same 3D half-space and friction law, and consists of two right-lateral, vertical strike-slip fault planes governed by linear slip weakening friction (Table S2).
Each fault is 30~km long along-strike ($x$-direction) and 20 km deep ($z$-direction), positioned parallel to each other, separated by a 3 km wide step-over ($y$-direction), with a 10~km along-strike overlap. The material properties and initial conditions are detailed in Table S2.

\begin{figure}[hptb]
    \centering
    \includegraphics[width=1.00\columnwidth]{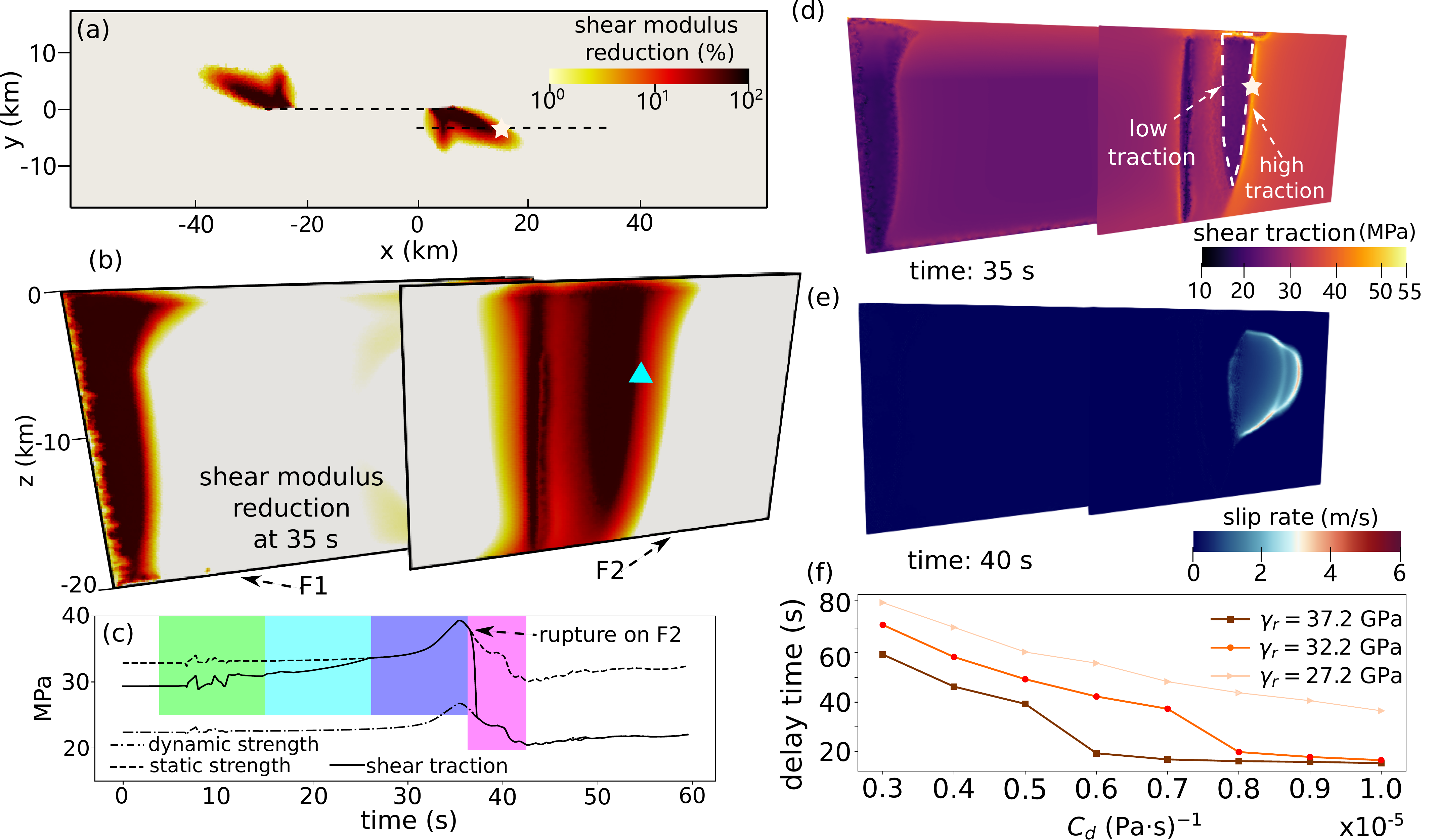}
    \caption{\small Delayed dynamic triggering across fault segments due to off-fault damage. (a) Shear modulus reduction distribution at 7.5 km depth, 35 s after rupture initiation, showing localized off-fault damage extending between faults F1 and F2. The white star shows the hypocenter of delayed triggered rupture on F2. (b) Close-up view of shear modulus distribution near the two faults, indicating the location of a receiver (cyan triangle) at (12.5, -3.0, -7.5) km. (c) Time series comparing shear traction (solid curve), static (dashed curve) and dynamic (dash-dotted curve) frictional shear strength at the receiver location indicated in (b) The black-dashed arrow marks the initiation of spontaneous rupture on fault F2. (d) Spatial distribution of shear traction on both faults at 35 s, with the hypocenter on F2 marked by a white star. 
    (e) Slip rate distribution at 40 s after fault F2 is dynamically delayed-triggered.
    (f) Variation in delay time between rupture initiation on fault F1 and the initiation on fault F2 as a function of the nonlinear modulus $\gamma_r$ and damage evolution coefficient $C_d$ in the CDB model (Eq. (2) in SI). Each marker represents delay times from an independent simulation; all parameters are provided in Table S2. We show simulations with varying $\gamma_r$ and $C_d$ in (f). Additional slip rate and shear traction distributions at intermediate time steps are presented in Fig. S5.}
    \label{Fig4:stepover}
\end{figure}

Figure \ref{Fig4:stepover} shows how co-seismic off-fault damage impacts delayed dynamic triggering between adjacent fault segments. Dynamic rupture nucleating on fault F1 induces localized zones of reduced shear modulus extending towards fault F2, producing a heterogeneous distribution of rock properties and stress between the faults (Figs.\ref{Fig4:stepover}a,b).
The initial rupture nucleation and propagation on fault F1 (Figs. S5a-1 and a-2) are similar to the elastic benchmark scenario and include supershear transition (Movie S3).
The dynamic and static stress perturbations are insufficient to trigger immediate rupture on fault F2.  
However, after the complete rupture of F1, localized zones of shear modulus reduction evolve from the end of F1 towards F2 (Fig. \ref{Fig4:stepover}a). This introduces significant heterogeneity in rock stiffness and stress distributions in the vicinity of F2 (Fig. \ref{Fig4:stepover}b). 

The dynamic damage and stress field evolution leading to delayed triggering of F2 involves four distinct phases (Fig. \ref{Fig4:stepover}c and Movie S4).
In phase I (green shading), the immediate dynamic and static stress perturbations from fault F1 reach fault F2 but remain below the fault's shear strength threshold.
During phase II (cyan shading), as the nonlinear off-fault damage zone around F1 expands towards F2, shear traction locally reduces within this damage zone (dashed white curve, Fig. \ref{Fig4:stepover}d). 
To balance the total frictional force on the fault, the neighboring rocks need to maintain higher traction. 
During phase III (blue shading), areas of increased shear traction imprint as three distinct transient high shear-traction fronts that slowly migrate ($<$0.1 km/s) alongside the evolving rock damage around fault F2 (Movie S4). These dynamic stresses do not cause fault slip (blue shading, Fig. S6a). However, these high shear-traction fronts are not aseismic but radiate seismic waves at frequencies below 0.03~Hz (non-zero $v_x$ with blue shading, Fig. S6b).

In phase IV (shaded pink),  the earthquake ``jumps'' to F2 with a considerable delay time. One of the damaged shear zones approaches F2, causing locally high enough shear stressing at one of the transient stress fronts to reach local fault shear strength across a critical area (white stars, Figs. \ref{Fig4:stepover}a,d), triggering delayed spontaneous dynamic rupture nucleation and propagation including a second supershear transition on fault F2 (Fig. \ref{Fig4:stepover}e).
Fault slip rapidly increases to the critical slip distance $D_c$ at this high shear-traction front (the dashed white arrow, Fig. \ref{Fig4:stepover}d) and the shear traction drops to its dynamic value (Fig. \ref{Fig4:stepover}c).
The rupture initiation on the second fault is delayed by $\sim$31 s after the complete rupture of the first fault and by $\sim$38 s after rupture initiation on F1. Hereafter, we refer to the time difference between the rupture onset on fault F1 and the rupture onset on fault F2 (shear traction dropping from the local static strength to the dynamic strength, Fig. \ref{Fig4:stepover}c) as the \textit{trigger delay time}.

Fig. \ref{Fig4:stepover}f summarizes results of our systematic investigation of how the delay time depends on key nonlinear parameters of the CDB model. For a fixed nonlinear modulus $\gamma_r$  of 37.2 GPa, we vary the damage evolution coefficient $C_d$ from 3.0 $\times 10^{-6}$ (Pa$\cdot$s)$^{-1}$ to 10.0 $\times 10^{-6}$ (Pa$\cdot$s)$^{-1}$. The trigger delay time increases from 14 s to 58 s when we use a smaller damage evolution coefficient $C_d$. Similarly, decreasing the nonlinear modulus $\gamma_r$ from 37.2 GPa to 27.2 GPa further prolongs the delay time from 58 s to 79 s. 
These results suggest an important role of co-seismic off-fault damage parameters in governing delayed dynamic triggering across fault systems.

\section{Discussion}
\label{sec:discussions}

We perform 3D dynamic rupture simulations in a model that incorporates off-fault behavior governed by a continuum damage breakage (CDB) model. We verify the numerical implementation by demonstrating that (1) simulated off-fault shear-band angles align with analytical CDB model solutions (Fig. \ref{Fig2:large damage, different angles, depths}), (2) energy components are conserved during dynamic rupture simulations (Fig. \ref{Fig3:energy budget}) and (3) localized off-fault damage patterns remain consistent with mesh refinement from 100 m to 25 m (Fig. S1).

The adopted CDB model employs two spatially continuous internal variables to characterize the pre- and post-failure states and mechanical behaviors of rocks. The gradual growth of crack density in intact rocks is represented with a damage variable $\alpha$ \citep{lyakhovsky1997distributed}. The rapid loss of stiffness at a critical value of $\alpha$ produces a dynamic brittle failure associated with a solid-granular phase transition and evolution of a breakage variable $B$, and the post-failure deformation of the granular is approximated with the breakage mechanics \citep{einav2007breakage1,lyakhovsky2014continuum}. With the two averaged internal variables over representative volumes, the CDB model avoids the explicit meshing of microscopic rock deficiencies in methods such as the finite-discrete element method \citep{okubo2019dynamics,mcbeck2022predicting}. This reduces the computational cost of the CDB model, enabling its application to 3D regional-scale earthquake simulations in this study. With such simplification, the CDB model still produces various important features of rupture dynamics including generation of fault damage zones with additional high-frequency radiation, and delayed dynamic triggering.

\subsection{High-frequency radiation from earthquake sources}
\label{subsec:high_f_radiation}

The simulated high-frequency radiation can explain detailed observations in laboratory experiments and in close proximity to earthquake ruptures. The high-frequency ($>$1 Hz) kinetic energy in off-fault regions is generated concurrently with the development of localized shear bands, which result from rapid solid-granular phase transitions leading to high damage in off-fault rocks behind the moving rupture front (Fig. \ref{Fig2:large damage, different angles, depths}, Movie S1). This is consistent with back-projection observations in laboratory stick-slip experiments on saw-cut granite samples by \citet{marty2019origin}.

Non-linear damage may be an important ingredient in physics-based simulations of high-frequency radiation \citep{shi2013rupture,withers2018ground}, which is usually modeled empirically \citep[e.g.,][]{boore1983stochastic} or stochastically \citep[e.g.,][]{graves2010broadband}. Better capturing of high-frequency observations may require to account for nonlinear site effects \citep{bonilla2011nonlinear,roten2016high,niu2025nlwave}, which contributes to more accurate ground motion simulations for seismic hazard analysis \citep{hanks1981character,chandramohan2016impact}.
For example, \citet{taufiqurrahman2022broadband} illustrate the potential of fully physics-based simulations in capturing broadband ground motions between 0.5 and 5 Hz during the 2016 M$_w$ 6.2 Amatrice earthquake using topography, viscoelastic attenuation and fault roughness. However, their 3D dynamic rupture simulations still underestimate the observed spectral amplitudes above 1 Hz.

Previous analytical and numerical results indicate that the high-frequency waves produced by rock damage are primarily isotropic \citep{ben2009seismic,lyakhovsky2016dynamic,zhao2024dynamic}. This is consistent with the results presented in Fig. \ref{Fig2-1:fnfp_ground_motions}), where we find that the ratios of the FN and FP components of high frequency radiation ($>$1~Hz) are close to 1.0, and depend only weakly on the azimuth angle from the epicenter. Such features were observed in recorded ground motions close to earthquake rupture zones \citep{graves2016kinematic,ben2024isotropic}. Additional observations consistent with isotropic damage-related radiation include inversions of near-fault seismograms for full source tensor source terms \citep{dufumier1997resolution,ross2015isotropic,cheng2021isotropic}, enhanced P/S amplitude ratios of high frequency waves \citep{satoh2002empirical,castro1991origin,castro2013potential} and elevated P/S ratios of the total radiated seismic energy \citep{garcia2004determination,kwiatek2013assessment}. 

Such observations cannot be explained with simulations assuming linear elastic off-fault materials. Our 3D dynamic rupture simulations with the CDB model can address this discrepancy by capturing co-seismic off-fault moduli reduction and their resulting isotropic high-frequency radiation patterns.

\subsection{Earthquake interaction with co-seismic off-fault damage}
\label{subsec:earthquake_interaction}

Our simulations reveal a novel mechanism in which co-seismic off-fault damage induces localized reductions in rock moduli, creating stress heterogeneities that enable delayed dynamic triggering across adjacent fault segments. 
The proposed new mechanism for delayed dynamic triggering arises from dynamic damage evolution and stress redistribution and consists of four distinct phases: (1) initial dynamic stress transfer; (2) expansion of localized non-linear damage zones, that radiate low-frequency seismic waves and cause local traction reduction; (3) formation of high shear-traction fronts around this damage zone; and (4) eventual delayed triggering, as rupture spontaneously nucleates on a secondary fault when localized shear traction reaches the frictional strength threshold across a critical area.
The delayed triggering depends primarily on the time required for the evolving damage zone to propagate and reach neighboring faults. 
As demonstrated in our 3D simulations, coseismic off-fault damage may effectively connect fault segments separated by distances of several kilometers, thereby facilitating rupture cascades in complex fault systems \citep{wesnousky2006predicting},
such as during the 2016 M$_w$ 7.8 Kaikoura earthquake \citep{bai2017two,ulrich2019dynamic}.
With variations in the damage evolution parameters, the modeled delay times range from several seconds up to tens of seconds (Fig.~\ref{Fig4:stepover}f).

In observations of large earthquake doublets (M$_w$$>$6), the trigger delay time ranges from a few to tens of hours \citep{hauksson19931992,ryder2012large,ross2019hierarchical,jia2023complex}. 
In our dynamic rupture simulations with the CDB model, the trigger delay time monotonously increases with smaller $\gamma_r$ and smaller $C_d$ (Fig. \ref{Fig4:stepover}f). This indicates that the delay time in the CDB model can be even longer than a few minutes with $C_d < 10^{-5}$ (Pa$\cdot$s)$^{-1}$ or $\gamma_r < 27.2$ GPa.
The non-linear modulus $\gamma_r$ depends on the two Lam\'{e} parameters $\lambda_0$, $\mu_0$ and the critical strain invariant ratio $\xi_0$ \citep{lyakhovsky2014continuum}, which is related to the internal friction angle of rocks \citep{griffiths1990failure}. 
For granite, the Lam\'{e} parameters typically range between 20 and 40 GPa \citep{ji2010lame}, and the internal friction angle varies between 25$^{\circ}$ and 45$^{\circ}$ \citep{wines2003estimates}, corresponding to a range of approximately 20–50 GPa.
Previous laboratory experiments on granite samples \citep{lyakhovsky2016dynamic} suggest a damage evolution coefficient $C_d$ within $10^{-9}$ to $10^{-7}$ (Pa$\cdot$s)$^{-1}$ at strain rates between $10^{-5}$ and $10^{-3}$ s$^{-1}$ \citep{lyakhovsky2016dynamic}.
In this study, the smallest $C_d$ is 3 $\times 10^{-6}$ (Pa$\cdot$s)$^{-1}$ (Fig. \ref{Fig4:stepover}f), but longer triggering delays, exceeding the tens of seconds to minutes range observed in our simulations, could occur under realistic rock conditions. 
Delayed triggering over longer time intervals that last days or more may be facilitated by additional evolution of rock damage through aftershocks and/or aseismic deformation.
To study delayed triggering on longer time scales will require developing a numerical implementation of the CDB model with adaptive explicit time step control \citep[e.g.,][]{uphoff2023discontinuous,yun2025effects} or an implicit time-stepping method \citep[e.g.,][]{pranger2020unstable}, instead of the explicit time-stepping in our implementation \citep{dumbser2006arbitrary,pelties2012three,wollherr2018off}

\section{Conclusions}
\label{sec:conclusions}
We present 3D dynamic rupture simulations incorporating nonlinear brittle off-fault damage to explore the interactions between seismic rupture, damage evolution, and seismic radiation. We analyze results associated with off-fault brittle damage during the gradual approach to brittle failure and during macroscopic dynamic rupture. 
Distinct damage regimes separated by the solid-to-granular transition emerge: smooth, distributed damage occurs under low damage conditions, transitioning to localized, mesh-independent shear bands upon reaching brittle failure. 

At low damage levels, off-fault damage dissipates significant energy, reducing rupture speed and inhibiting transitions to supershear rupture propagation. Damage accumulation is locally reduced at the supershear transition zone because of the more compressive strain field. 
In addition, the generated damage zones exhibit depth-dependent variations, widening significantly toward the Earth's surface even under uniform background stress, aligning with field observations.

When off-fault damage exceeds the threshold of brittle failure, shear bands evolve that align systematically with the background stress state and are consistent with analytical predictions. Co-seismic damage generates pronounced high-frequency seismic radiation above 1 Hz, producing near-isotropic fault-normal and fault-parallel high-frequency ground motions, consistent with observations.

We identify a novel mechanism for delayed dynamic triggering in multi-fault systems, driven by localized reductions in elastic moduli and associated static stress heterogeneity around tensile fault step-overs.  With the combined effects of damage-induced high-frequency radiation and off-fault energy dissipation, we find that the off-fault damage only alters the total kinetic energy by less than 1\%. This suggests negligible effects on the dynamic stress perturbations of the neighboring faults. In contrast, the static stress field is more strongly influenced by rock damage and enhances the fault triggering, with a delay time, in the tensile stepover configuration. This mechanism promotes rupture cascading across fault segments, with the delay time strongly influenced by the damage evolution coefficient ($C_d$) and nonlinear modulus ($\gamma_r$). Smaller values of $C_d$ or $\gamma_r$ can prolong the delay time from a few seconds to a few minutes. 

Our findings offer a physics-based explanation for enhanced high-frequency seismic radiation and delayed rupture triggering, advancing our understanding of earthquake processes, seismic radiation characteristics, and complex fault interactions.
This work also provides a unique, openly available tool that can model how co-seismically evolved fault zone damage changes earthquake source mechanisms and may provide more realistic high-frequency ground motions in three-dimensional earthquake simulations.

\acknowledgments

We thank Heiner Igel and Vladimir Lyakhovsky for discussions about the setup of the model. We also thank Sebastian Wolf for his support with the implementation of the CDB model in SeisSol.
ZN and AAG were supported by the European Union's Horizon 2020 research and innovation programme under the Marie-Sklodowska-Curie grant agreement No. 955515 – SPIN ITN (www.spin-itn.eu) and the Inno4scale project, which is funded by the European High Performance Computing Joint Undertaking (JU) under Grant Agreement No. 101118139. The JU receives support from the European Union's Horizon Europe Programme. 
AAG acknowledges additional support from the National Science Foundation (grant nos. EAR-2225286, EAR-2121568, OAC-2311208 and OAC-2311206) and the National Aeronautics and Space Administration (grant no. 80NSSC20K0495). The work of YBZ on analysis and interpretation was supported by the U. S. Department of Energy (Award DE‐SC0016520). 
The authors acknowledge the Gauss Centre for Supercomputing e.V. (www.gauss-centre.eu) for providing computing time on the supercomputer SuperMUC-NG at the Leibniz Supercomputing Centre (www.lrz.de) in project pn49ha. Additional computing resources were provided by the Institute of Geophysics of LMU Munich \citep{oeser2006cluster}.

\section*{Open Research}

The source code of SeisSol with the continuum damage breakage model implementation is available as open-source software from \citet{uphoff_2024_14051105} under the branch \texttt{damaged-material-nonlinear-drCDBM}. The model setup, simulation outputs, and post-processing scripts to reproduce all figures are available from \citet{niu_2025_15034486}.

\newpage

\appendix

\section{Mesh-independent damage}
\label{app:numerics}
Achieving mesh-independence in numerical simulations of nonlinear continuum damage models is crucial to ensure physically meaningful and reliable model results \citep[e.g.,][]{riesselmann2023simple}. 
We demonstrate that our implementation of the CDB model within the discontinuous Galerkin framework produces mesh-independent off-fault damage patterns across element sizes ranging from 100 m to 25 m (Fig.~S1).

Mesh independent continuum damage modeling typically relies on numerical relaxation \citep{gurses2011evolving} or spatial regularization techniques using damage gradients \citep{peerlings1996gradient,lyakhovsky2011non}. In our CDB-DG implementation, we achieve mesh-independent behavior without explicit regularization (see Eq. (2) of the SI).
This mesh-independence is due primarily to numerical diffusion introduced by the Rusanov flux \citep{rusanov1961calculation,leveque2002finite}, as detailed in \citet{niu2025nlwave}.
Similarly mesh-independent results have been achieved for for nonlinear hyperelasticity with material failure using a DG method with a diffusive subcell finite-volume limiter \citep{tavelli2020space}.

Mesh-independence simplifies the requirements for incorporating realistic co-seismic off-fault damage in regional-scale earthquake simulations. For example, in our simulations, we achieve accurate high-frequency ground motions up to 4 Hz within 10 km of the source using $p=1$ polynomial basis functions and mesh elements as large as 100 m near the fault, coarsening to 300 m at 10 km distance and further to 5 km at greater distances. This results in a mesh with $\sim$5.5 million tetrahedral elements. The simulation for 10 s takes $\sim$2560 CPU hours on SuperMUC-NG (phase 1) with Intel Xeon Platinum 8174 processors.

\bibliography{references}

\end{document}